\documentclass[preprint,12pt]{elsarticle}
\usepackage{epsfig}
\usepackage{color}
\usepackage{amsmath}
\usepackage{amssymb}
\usepackage{dsfont}
\usepackage{cancel}
\usepackage[normalem]{ulem}

\def\beq{\begin{equation}}
\def\eeq{\end{equation}}
\def\bea{\begin{eqnarray}}
\def\eea{\end{eqnarray}}
\def\sea{\nonumber \\&&}
\def\lla{\left\langle}
\def\rra{\right\rangle}

\def\zb{\beta}
\def\zc{\gamma}
\def\ssc{\scriptscriptstyle}
\def\lsim{\mathrel{\raise.3ex\hbox{$<$\kern-.75em\lower1ex\hbox{$\sim$}}} }
\def\gsim{\mathrel{\raise.3ex\hbox{$>$\kern-.75em\lower1ex\hbox{$\sim$}}} }

\usepackage{graphicx,accents}

\makeatletter
\DeclareRobustCommand{\cev}[1]{%
  \mathpalette\do@cev{#1}%
}
\newcommand{\do@cev}[2]{%
  \fix@cev{#1}{+}%
  \reflectbox{$\m@th#1\vec{\reflectbox{$\fix@cev{#1}{-}\m@th#1#2\fix@cev{#1}{+}$}}$}%
  \fix@cev{#1}{-}%
}
\newcommand{\fix@cev}[2]{%
  \ifx#1\displaystyle
    \mkern#23mu
  \else
    \ifx#1\textstyle
      \mkern#23mu
    \else
      \ifx#1\scriptstyle
        \mkern#22mu
      \else
        \mkern#22mu
      \fi
    \fi
  \fi
}

\makeatother

\begin{document}

\begin{frontmatter}

\title{The Noncommutative Values of Quantum Observables
\vspace*{.3in} }
\author{Otto C. W. Kong}
\ead{otto@phy.ncu.edu.tw} 
\cortext[cor1]{Corresponding author}
\author{Wei-Yin Liu}

\address{Department of Physics and 
Center for High Energy and High Field Physics\\
National Central University, Chung-Li 32054, Taiwan
}

\begin{abstract}
We discuss the notion of representing the values of 
physical quantities by the real numbers, and its limits to
describe the nature to be understood in the relation to our 
appreciation that the quantum theory is a better theory
of natural phenomena than its classical analog. Getting
from the algebra of physical observables to their values
for a fixed state is, at least for classical physics, really a
homomorphic map from the algebra into the real number
algebra. The limitation of the latter to represent the values
of quantum observables with noncommutative algebraic
relation is obvious. We introduce and discuss the idea of
the noncommutative values of quantum observables and
its feasibility, arguing that at least in terms of the
representation of such a value as an infinite set of
complex numbers, the idea makes reasonable sense
theoretically as well as practically.
\end{abstract}


\end{frontmatter}

\section{Introduction}
Newton presented his theory of mechanics with the hypothetical
notion of a point particle with an unambiguous position
(the point) in the physical space of assumed three dimensional
Euclidean geometry as essentially a product of the real number 
lines. Note that kind of geometry was the only mathematical
model for the space as a continuum in the mathematics of 
Newton's time. The three independent coordinates of a particle
then give the basic observables the values of which are given 
by the real numbers. Classical physics is then a real number physics.
Physical quantities or observables, of the particle in the case, are 
modeled by the real valued variables. However, in order to allow the 
theory to work in all inertial frames, instead of only in an absolute 
frame of reference, with respect to which we measure all the motion, 
the basic independent variables have to be extended to include the 
velocity or momentum components. We have the full set of six 
real coordinates of a point in the phase space giving a unique
state of the particle. The phase space is then also essentially 
modeled by the Euclidean geometry, though the idea of a metric
or distance between two states is not considered to be of any
use.  The picture of a particle as occupying a fixed position in
physical space is an intuitive one, a desirable if not an absolutely
necessary feature of the theory of particle dynamics (versus
field theories) that we have been told we have to give up in
quantum mechanics. A key feature of our study of quantum 
spacetime is to restore that. A quantum particle sure cannot
have a fixed position in the Newtonian model of the physical
space. That may rather be taken as an indication that the 
latter model fails the intuitive notion of a model for the
physical space and demands efforts to find a good enough
model that works. In fact, the only physical notion of the
physical space in a theory of particle dynamics is the 
collection of all possible positions of a free particle. Actually,
the model for the physical space is such an integral part of 
the theory  that changing the theory assuming the
model to be unchanged may not be really sensible. 

The most important point of departure from classical 
physics in quantum mechanics is the realization that 
quantum observables, the necessary model description 
of real physical observables as seen at high enough precision 
beyond the classical domain, in general do not commute. 
In particular, the conjugate pairs of position and momentum 
observables each is an non-commuting pair.  The position 
observables, however, commute among themselves. The 
latter might be the reason why most physicists believe it 
is still fine to keep the Newtonian model for the physical 
space. But then the quantum phase space has been realized 
to be very different from the Cartesian product of the three 
dimensional Euclidean spaces for the position and the 
momentum. It is an infinite dimensional Hilbert space 
$\mathcal{H}$, or better taken as the projective Hilbert 
space $\mathcal{P}$ (or a $U(1)$ principal bundle of it). With 
time came the appreciation of the latter as a symplectic 
manifold with the Schr\"odinger equation giving an infinite 
number of pairs of Hamilton equation of motion on it (see 
our companion paper \cite{078} and references therein). That 
begs the questions why we cannot take a certain Lagrangian
submanifold of it, which should correspond to the configuration 
space for a free particle, as a model for the physical space, 
and how to reconcile the fact of there being only the three
pairs of position and momentum observables with 
the infinite dimensional phase space which could be taken 
as having an infinite number of pairs of position and momentum 
coordinates. Answer to the first question has been presented in 
Refs.\cite{066,070}. Those position and momentum 
coordinates are like the space and time coordinates of the 
Minkowski spacetime, separable only in the Newtonian 
approximation. In the quantum setting, there cannot be an
independent notion of a configuration space, rather it can only
be seen as part of the phase space. The phase space should be 
taken as the proper model for a sensible notion of something
like the configuration space, hence (that for a free particle)
should be taken as the model for the physical space.
The current study, together with Ref.\cite{078}, is to 
address the second question. The former paper looks into
how to think about the noncommutative geometric picture 
of the otherwise well known infinite dimensional K\"ahler
manifold, to the point of illustrating a quite explicit 
description of a transformation between the infinite 
number of complex number coordinates to the six 
noncommutative coordinates as complex combinations
of the pairs of position and momentum operators. 
Matching of the full differential and symplectic structures
are presented in that language. The current article focus
on the new notion of noncommutative value for a (quantum)
observable, mostly through a description of such a value
in terms of an infinite set of real numbers. The results
establish the perspective that the information content 
of the set of the six noncommutative coordinates is the 
same as the of the infinite set of real/complex number  
coordinates. We take it as a pure theoretical/mathematical 
question first, and discuss the issues of the practical 
implementation of the results only at the end. A complete
real number coordinate summary version of the key
results here using simple mathematics,
by taking the Hilbert space as phase space, with emphasis
on the integrated perspectives for an intuitive picture of
quantum/noncommutative physics is presented in 
Ref.\cite{081} for a more general readership.

\section{The Noncommutative Values of Observables}
Let us focus first on the observables. Like the classical
case, we think about all quantum observables as functions
of the basic set of independent observables, the position
and momentum ones. Each such classical observable is 
then a real variable in itself, while in the quantum case it
is an element of a noncommutative observable algebra 
which has been modeled by the operators on $\mathcal{H}$.
What about the values for such operators on a state? In
 the classical case, the value of an observable $a(p^i,x^i)$ 
for a state $(p^i_o,x^i_o)$ is the real number $a(p^i_o,x^i_o)$. 
Considering the values for all observables in the algebra
at the same time, we have an evaluation map for each
state that maps all the variables $a(p^i,x^i)$ to the real 
number values of $a(p^i_o,x^i_o)$. Such a map has to be 
a homomorphism between the observable algebra and 
the real numbers as an algebra. That is to say the algebraic
relationship with a number of observables has to be
respected by their values for any state. The evaluation map 
given by taking the real number values of the observables
as functions of the states is obviously such a homomorphism. 
To push that reasoning to the algebra of quantum 
observables, we need  a picture of the values for 
$\hat{x}^i$ and $\hat{p}^i$ satisfying also the commutation 
relation $[ \hat{x}^i, \hat{p}^j] = i\hbar \delta^{ij}$ as the 
image of an evaluation map for a state here denoted by 
$[\phi]$. $\left|\phi\rra$ generally denotes a state vector 
in the Hilbert space. An actually physical state of course
corresponds to the ray determined by any such vector
related to one another by a constant multiple. Each
physical state would be seen as essentially an evaluation 
map, as in the classical case, which fixes the values of
all observables, the values they have for the state.  
However, the algebra obtained as the image of such 
a map has then to be a noncommutative algebra, for which a
subalgebra of the real number algebra certainly cannot do. 
That in itself does not necessarily imply that we cannot have
such maps, thus to have the fixed values for the quantum
observables for a state. We cannot have the fixed real number
values. So long as we are willing to consider using some
noncommutative algebraic system to represent those values,
it is plausible to have the definite values. The key purpose of 
this article is to give an explicit description of the kind of algebraic 
model, one we may call an algebra of noncommutative numbers. 
Actually, an isomorphic description of the observable algebra with 
reference to a definite state has been presented in a 1996 Ph.D. 
dissertation \cite{Sd},  though the author(s) did not seem to 
have the idea of that being the algebra of noncommutative 
(number) values for the observables. What the author(s) called 
{\em the algebra of symmetry data, with a particular set of 
infinite number of complex numbers as an element, is really the 
candidate for the algebra of noncommutative values, may be
called the noncommutative number values, of the observables 
in a quantum theory. That is the key proposition in this article. } 
We aim at presenting below a more direct picture of that algebraic
story and start to seek a better understanding and depiction of 
physics in terms of such noncommutative number values of the 
observables.
 
To prepare for the appreciation of the mathematics, we first 
want to note that an operator has really the information content 
of an infinite number of real/complex numbers. Taking an 
orthonormal basis for $\mathcal{H}$, for example the eigenstates 
of the `three dimensional' harmonic oscillator, an operator
is completely characterized by the infinite set of matrix 
elements. However, the picture is independent of any state. 
The symmetry data on a state for each noncommutative 
observable/variable is exactly a state specific set of infinite 
number of complex numbers characterizing the observable
hence giving the noncommutative value of it, and such sets 
for the various observables have a noncommutative product 
that gives the set for the value of the product observable from
the values, {\em i.e.} the two sets of complex numbers,
for the two observables composing it. Of course such a set of
complex numbers is only one mathematical way to represent the
noncommutative value as an element of a noncommutative
algebra. It is the way that  gives it a more concrete realization 
for those familiar with the real/complex number values. 
Yet, the real numbers themselves as mathematical symbols
are not really fundamentally any less abstract than elements
of other algebras. 
 
To sketch the story of the symmetry data in Ref.\cite{Sd}
and present our formulation, we have to first introduce
two isomorphic descriptions of the observable algebra,
which are more or less given in the reference, but actually
have their origin in a very important earlier paper by Cirelli,
Mani\`a, and Pizzocchero \cite{Cps} to which we owe much
for our current understanding of the geometric structures
behind quantum mechanics. Ref.\cite{Sd} takes it quite a bit 
further, clarifying the geometric relations between $\mathcal{H}$ 
and $\mathcal{P}$ from the perspective of quantum mechanics 
and presenting the so-called symmetry data as a state 
dependent algebra isomorphic to the observable algebra.
Apart from advancing the idea of the latter as candidates for the 
noncommutative values of observables, we will also present 
more explicit results in terms of standard complex coordinates 
to give readers easy comprehension of the key mathematics. 
Some details are left to Ref.\cite{078} though. The 
algebraic isomorphisms are
$H: \zb \to H_{\!\ssc\zb}(z^n,\bar{z}^n)$
and $f:  \zb \to f_{\!\ssc\zb}(z^n,\bar{z}^n)$
with
\bea
H_{\!\ssc\zb}(z^n,\bar{z}^n) =\frac{1}{2\hbar} \lla \phi | \zb | \phi \rra
  = \frac{1}{2\hbar} \sum \bar{z}^m z^n \lla m | \zb | n \rra \;,
\eea
and
\bea
f_{\!\ssc\zb}(z^n,\bar{z}^n) = \frac{2\hbar}{|z|^2} H_{\!\ssc\zb}(z^n,\bar{z}^n) 
  = \frac{1}{|z|^2} \sum \bar{z}^m z^n \lla m | \zb | n \rra \;,
\eea
for an operator $\zb$, of which we prefer to think about as 
a function of the coordinate operators $\zb(\hat{p}^i, \hat{x}^i)$,
with state vector $\left|\phi\rra = \sum  z^n \left| n \rra$; 
$\left| n \rra$ with $n$ from 0 to $\infty$ denote an orthonormal 
basis of $\mathcal{H}$ and $z^n$ (with the index $n$) the complex 
coordinates of a point in $\mathcal{H}$ as a K\"ahler manifold;
$|z|$ is then the norm of the state vector, or $|z|^2=z_n\bar{z}^n$,
Einstein summation assumed as throughout the paper.
$f_{\!\ssc\zb}$ can be taken as functions on $\mathcal{H}$ or 
functions on $\mathcal{P}$ here in homogeneous coordinates. 
Note that they are independent of the normalization of 
$\left|\phi\rra$. The algebras of $H_{\!\ssc\zb}$ or 
$f_{\!\ssc\zb}$ functions of course each need to have 
a noncommutative product to match that of the operator 
product. They are the K\"ahler products given by
\bea
H_{\!\ssc\zb} \star_{\!\ssc K} H_{\!\ssc\zc}
= \hbar \, \partial_{{m}}  H_{\!\ssc\zb}  \,G^{{m} \bar{{n}}} \partial_{\bar{{n}}} H_{\!\ssc\zc} \;,
\eea
and
\bea
 f_{\!\ssc\zb} \star_{\kappa} f_{\!\ssc\zc} =  f_{\!\ssc\zb} f_{\!\ssc\zc} 
   + \hbar \, \partial_m f_{\!\ssc\zb} \, 
      \tilde{g}^{m\bar n}\partial_{\bar n} f_{\!\zc} \;,
\eea
where $G^{{m} \bar{{n}}}= 2 \delta^{m\bar{n}}$ is the 
inverse metric of $\mathcal{H}$ and $\tilde{g}^{m\bar n}$ 
that of $\mathcal{P}$. The metric of $\mathcal{P}$ is the 
standard Fubini-Study metric. When expressed in terms 
of the homogeneous coordinates, the latter is degenerate. 
However, we can still have $\tilde{g}^{m\bar n}$ from
Killing reduction of $\mathcal{H}-\{0\}$ \cite{Sd,078}, given as  
$\frac{1}{\hbar} \left( |z|^2\delta^{m\bar{n}}   -  {z}^{{m}} \bar{z}^{{n}} \right)$ 
which can be used in the K\"ahler product calculation.  
We can also think about $\mathcal{H}-\{0\}$ as a fiber 
bundle with $\mathcal{P}$ as the base manifold and the 
degenerate metric independent of the fiber 
coordinates. The K\"ahler product may otherwise be 
given for the $f_{\!\ssc\zb}$ functions expressed in terms 
of a set of affine coordinates, for example 
$w^n=\frac{z^n}{z^{\ssc 0}}$ for $n \ne 0$ with the 
non-degenerate Fubini-Study metric  \cite{078}.  In that case,
we have  $f_{\!\zb} \star_{\kappa} f_{\!\zc} = f_{\!\zb} f_{\!\zc} 
   + \hbar \, \partial_m f_{\!\zb} \, {g}^{m\bar n}\partial_{\bar n} f_{\!\zc}$,
 with ${g}^{m\bar n} =\frac{1}{\hbar} ({1+|w|^2}) 
     \left( \delta^{m\bar{n}} +{w^m \bar{w}^n}  \right)$.
We have $H_{\!\ssc\zb\zc}=  H_{\!\ssc\zb} \star_{\!\ssc K} H_{\!\ssc\zc}$
and $f_{\!\ssc\zb\zc}=  f_{\!\ssc\zb} \star_{\kappa} f_{\!\ssc\zc}$
which can be easily verified explicitly in terms of generic
matrix elements for the operators based on the coordinate.

The symmetry data for an operator $\zb$ at a point 
$[\phi]$ on $\mathcal{P}$ is given in Ref.\cite{Sd} as 
a triplet $(f_{\!\ssc\zb}, X_{\!\ssc\zb}, K_{\!\ssc\zb})$ with 
$X_{\!\ssc\zb}$ being a covector of the Hamiltonian vector 
field for $f_{\!\ssc\zb}$,  and $K_{\!\ssc\zb}$ a 2-form as 
the covariant derivative of $X_{\!\ssc\zb}$, all taken with 
values at $[\phi]$. Explicit expressions for the symmetry 
data for the Poisson bracket and Riemann bracket of two 
$f_{\!\zb}$ functions, corresponding to commutator and 
anticommutator of the operators \cite{078,Sd,Cps}, in terms 
of the two sets of  $(f_{\!\ssc\zb}, X_{\!\ssc\zb}, K_{\!\ssc\zb})$ 
for the operators, are presented \cite{Sd}.  We present here 
below our reformulation of the picture which we believe 
makes more transparent  all otherwise hidden in the 
abstract mathematics. And we give directly the product
of the algebra of symmetry data at $[\phi]$ in terms of 
the K\"ahler product for the original $f_{\!\ssc\zb}$ 
functions. The K\"ahler product among the $f_{\!\ssc\zb}$ 
functions represents the operator product, to which 
the Poisson and Riemann brackets  are simply the 
antisymmetric and symmetric parts. 

As usual, mathematics behind quantum mechanics is easier 
to present on $\mathcal{H}$. Let us first look at the symmetry 
data for the $H_{\!\ssc\zb}$ functions. For each operator
$\zb$, it is given as $(H_{\!\ssc\zb}, \widetilde{X}_{\!\ssc\zb_n}, 
  \widetilde{X}_{\!\ssc\zb_{\bar{n}}},
  \widetilde{K}_{\!\ssc\zb_{m\bar{n}}}\!\!
  =\!{\nabla}_{\!\!{{m}}} \widetilde{X}_{\!\ssc\zb_{\bar{n}}}$), 
where $\widetilde{X}_{\!\ssc\zb}$ is the Hamiltonian vector 
field of $H_{\!\ssc\zb}$. Explicitly, 
$\widetilde{X}_{\!\ssc\zb_n}=i \partial_n H_{\!\ssc\zb}$,
$ \widetilde{X}_{\!\ssc\zb_{\bar{n}}} =-i \partial_{\bar{n}} H_{\!\ssc\zb}$.
Here, we have each fixed $\left| \phi_{\ssc 0} \rra$ gives an 
evaluation map $[\left| \phi_{\ssc 0} \rra]$ that determines 
the complex number value for each expression in 
 $(H_{\!\ssc\zb}, \widetilde{X}_{\!\ssc\zb_n}, 
  \widetilde{X}_{\!\ssc\zb_{\bar{n}}},
  \widetilde{K}_{\!\ssc\zb_{m\bar{n}}}\!\!
  =\!{\nabla}_{\!\!{{m}}} \widetilde{X}_{\!\ssc\zb_{\bar{n}}}$)
for all elements $\zb$ of the observable algebra. Note that 
each expression, {\em e.g.} $\widetilde{X}_{\!\ssc\zb_{\bar{n}}}$
is a function of the coordinates which have fixed real number
values for $\left| \phi_{\ssc 0} \rra$. The map $[\left| \phi_{\ssc 0} \rra]$ 
evaluates, or assigns a noncommutative value, to each specific 
observable $\zb$ as given here as the particular set of infinite 
number of complex numbers which is really an element of 
a noncommutative algebra. We are on our way to give the
product within the latter algebra, one that gives the 
noncommutative value of $\zb\zc$ from those of $\zb$
and $\zc$, for the same state vector $\left| \phi_{\ssc 0} \rra$.
Note that $\zb(\hat{p}^i, \hat{x}^i)$ 
are in general `complex functions', meaning  non-Hermitian 
operators, essentially complex linear combinations of Hermitian
operators, making up the observable algebra as a $C^*$-algebra. 
Therefore, $H_{\!\ssc\zb}$ is in general complex too. The real 
ones correspond to the Hermitian operators. The same holds 
for the $f_{\!\ssc\zb}$ functions.\footnote{
For the preferred formulation within our 
perspective of quantum relativity \cite{070}, we have
representation of the coordinate operators $\hat{p}_i$
and $\hat{x}_i$ as ${p}_i\star=p_i - i\partial_{\!x^i}$ and 
${x}^i\star =x_i +i \partial_{\!p^i}$, respectively, on the
space of coherent state wavefunctions $\phi(p^i,x^i)$, 
with $\zb(\hat{p}^i, \hat{x}^i)$ given by 
$\zb(p^i,x^i)\star$, with the $\star$ as the one in the Moyal
star product (here in $\hbar=2$ units). The latter is Hermitian 
for a real function $\zb(p^i,x^i)$ and non-Hermitian for a 
complex one. Actually, the operator product 
$\zb(p^i,x^i)\!\star~\zc(p^i,x^i)\star$ is then exactly
$[\zb(p^i,x^i)\star \zc(p^i,x^i)]\star$.}
We have, for the K\"ahler manifold with $H_{\!\ssc\zb}$
functions generating Hamiltonian flows as isometries, 
${\widetilde{K}}_{\!\ssc\zb_{n\bar{m}}} = - {\widetilde{K}}_{\!\ssc\zb_{\bar{m}n}} 
  =   -i  {\partial}_{{{n}}}  {\partial}_{\bar{{m}}}   H_{\!\ssc\zb}$
and  ${\widetilde{K}}_{\!\ssc\zb_{m{n}}}  
= {\widetilde{K}}_{\!\ssc\zb_{\bar{m}\bar{n}}} =0$.
The K\"ahler product gives
\bea &&
H_{\!\ssc\zb\zc} = {\hbar} \widetilde{X}_{\!\ssc\zb_m} G^{m\bar{n}}
    \widetilde{X}_{\!\ssc\zc_{\bar{n}}} 
={2\hbar} \sum_n \widetilde{X}_{\!\ssc\zb _{n}} \widetilde{X}_{\!\ssc\zc_{\bar{n}}}\;,
\sea
\widetilde{X}_{\!\ssc\zb\zc_{\,{n}}}
  = 2i\hbar  \sum_m  \widetilde{X}_{\!\ssc\zb _{m}} \widetilde{K}_{\!\ssc\zc_{n\bar{m}}} \;,
\sea
\widetilde{X}_{\!\ssc\zb\zc_{\,\bar{n}}} 
= 2i\hbar  \sum_m  \widetilde{K}_{\!\ssc\zb _{m\bar{n}}} \widetilde{X}_{\!\ssc\zc_{\bar{m}}} \;,
\sea
\widetilde{K}_{\!\ssc\zb\zc _{\,m\bar{n}}}
   = {2i\hbar}  \sum_l \widetilde{K}_{\!\ssc\zb _{l\bar{n}}} 
        \widetilde{K}_{\!\ssc\zc _{m\bar{l}}}\;.
\label{SDp-h}\eea
The set $\{ \widetilde{X}_{\!\ssc\zb_n}, \widetilde{X}_{\!\ssc\zb_{\bar{n}}},
  \widetilde{K}_{\!\ssc\zb_{m\bar{n}}} \}$
is really the set of all independent nonzero derivatives of $H_{\!\ssc\zb}$;
$\widetilde{X}_{\!\ssc\zb_{\bar{n}}}$ being the conjugate of
$\widetilde{X}_{\!\ssc\zb_{\bar{n}}}$ only for the Hermitian 
$\zb$. The second order derivatives give the matrix elements,
 $\widetilde{K}_{\!\ssc\zb_{m\bar{n}}}  
= - \frac{i}{2\hbar} \lla n | \zb | m \rra$.  
They are not dependent on the state at all, and the 
$\widetilde{K}_{\!\ssc\zb\zc _{\,m\bar{n}}}$ 
expression above simply gives the matrix elements of $\zb\zc$ 
in terms of the matrix elements of $\zb$ and $\zc$. The latter 
alone gives an isomorphic description of the observable algebra.  
The first order derivatives are, however, linear functions of 
the state coordinates, in particular
$\widetilde{X}_{\!\ssc\zb_n} = \frac{i}{2\hbar} 
   \sum_m \bar{z}^m\lla m | \zb | n \rra$.
Putting together the values of the first and second 
derivatives, say $\widetilde{X}_{\!\ssc\zb_n}$
and  $\widetilde{K}_{\!\ssc\zb _{m\bar{n}}}$, 
may allow us to obtain the values of the coordinates 
and hence determine the state vector $\left|\phi \rra$
up to an overall phase on which the $H_{\!\ssc\zb}$
functions have no dependence. That is
feasible for any invertible operator in itself. 

We are more interested in results for the $f_{\!\ssc\zb}$ 
functions. Unlike the Hilbert space function $H_{\!\ssc\zb}$,
$f_{\!\ssc\zb}$ as a function on the projective Hilbert space
has (affine) coordinates $w^n$. However, expressions in
terms of the $z^n$ as homogeneous coordinates are still
useful. Hence we present both. Note that though a set of
affine coordinates can be easily given from a set of 
homogeneous coordinates, the correspondence is not
unique and no set of affine coordinates in itself can cover 
the whole projective Hilbert space as a geometric 
manifold. As an  $f_{\!\ssc\zb}$ function is really
independent of the normalization of $\left|\phi\rra$,
using the $z^n$ coordinates for a $\left|\phi_{\ssc 0}\rra$
to determined an evaluation map $[\left|\phi_{\ssc 0}\rra]$
has only an overall complex phase ambiguity as a common
factor to all $z^n$. Otherwise, it corresponds uniquely
to a physical state as a point in the projective Hilbert 
space specified by the $w^n$ coordinates. The definite
evaluation map $[\phi_{\ssc 0}]$ for such a physical state
is free from such ambiguity, and evaluates by applying
fixed $w^n$ coordinates.

Using the $z^n$ coordinates with 
$\tilde{\omega}^{m\bar{n}} =-i\tilde{g}^{m\bar{n}}$,  we have
\bea &&
{\tilde{X}}_{\!\ssc\zb_n} = i \partial_n f_{\!\ssc\zb}
 =   \frac{- i}{|z|^2}  \Big[ f_{\!\ssc\zb} \bar{z}_n
   - \sum_m \bar{z}^m\lla m | \zb | n \rra \Big] \;,
\sea 
{\tilde{X}}_{\!\ssc\zb \bar{n}} =   - i \partial_{\bar{n}} f_{\!\ssc\zb}
=  \frac{i}{|z|^2}  \Big[   f_{\!\ssc\zb} {z}_n
   -  \sum_m {z}^m\lla n | \zb | m \rra  \Big]\;,
\sea
{\tilde{K}}_{\!\ssc\zb_{m\bar{n}}} = - {\tilde{K}}_{\!\ssc\zb_{\bar{n}m}} 
= - i \partial_{{m}} \partial_{\bar{n}} f_{\!\ssc\zb}
\sea\qquad
  = \frac{i}{|z|^2}  \big[ f_{\!\ssc\zb} \, \delta_{m\bar{n}}  
        +i {\bar{z}_m} {\tilde{X}}_{\!\ssc\zb_{\bar{n}}}
        -i {{z}_n}  {\tilde{X}}_{\!\ssc\zb_m} 
        -  \lla n | \zb | m \rra \big] \;.
\label{SD-z}\eea
Using the $w^n$ coordinates (no $n=0$) with
${\omega}^{m\bar{n}}=-i{g}^{m\bar{n}}$, we have 
\bea &&
{{X}}_{\!\ssc\zb_n} = i \partial_n f_{\!\ssc\zb}
 = \frac{-i}{(1+|w|^2)}  \Big[ f_{\!\ssc\zb} \bar{w}_n
   - \sum_m \bar{w}^m\lla m | \zb | n \rra \Big]\;,
\sea 
{{X}}_{\!\ssc\zb \bar{n}} =   - i \partial_{\bar{n}} f_{\!\ssc\zb}
=  \frac{i}{(1+|w|^2)} \Big[ f_{\!\ssc\zb} {w_n}
     -  \sum_m {w}^m\lla n | \zb | m \rra  \Big]\;,
\sea
{{K}}_{\!\ssc\zb_{m\bar{n}}} = - {{K}}_{\!\ssc\zb_{\bar{n}m}} 
= - i \partial_{{m}} \partial_{\bar{n}} f_{\!\ssc\zb}
\sea\qquad
  = \frac{i}{(1+|w|^2)} \big[ f_{\!\ssc\zb} \, \delta_{m\bar{n}}
        + i   {\bar{w}_m} {{X}}_{\!\ssc\zb_{\bar{n}}}
         - i  {w}_n  {{X}}_{\!\ssc\zb_m}  - \lla n | \zb | m \rra \big] \;,
\label{SD-w}\eea
where the index for the components, as the coordinates, 
runs from one but the summations includes $m=0$ with 
$w^{\ssc 0}=\bar{w}^{\ssc 0}=1$. The similar form of the
two sets of the results is deceiving. Apart from the different
coordinate derivatives, denoted by the same $\partial_n$
symbol for simplicity (with a pair less in the second case), 
the Hamiltonian vector field ${\tilde{X}}_{\!\ssc\zb}$ is 
obtained from the symplectic form $-i\tilde{g}^{m\bar{n}}$
while ${{X}}_{\!\ssc\zb}$ from the symplectic form 
$-i{g}^{m\bar{n}}$, though for the same $f_{\!\ssc\zb}$
function. Actually, ${\tilde{X}}_{\!\ssc\zb}$ and  
$-i\tilde{g}^{m\bar{n}}$ are, as tensors, the horizontal
lifts of ${{X}}_{\!\ssc\zb}$ and $-i{g}^{m\bar{n}}$,
respectively.  Their covectors are identical though, {\em i.e.} 
$\tilde{X}_{\!\ssc\zb_n} dz^n = {X}_{\!\ssc\zb_n} dw^n$
(sum without $n=0$ for the right hand side), because 
the ${{X}}_{\!\ssc\zb}$ covector is horizontal  \cite{078}.
Applying the K\"ahler product, we have
\bea &&
f_{\!\ssc\zb\zc}  \!
  = f_{\!\ssc\zb}  f_{\!\ssc\zc}  
   + |z|^2     \sum_n \tilde{X}_{\!\ssc\zb_n} \tilde{X}_{\!\ssc\zc_{\bar{n}}} \;,
\sea
\tilde{X}_{\!\ssc\zb\zc_{\,{n}}}  \!
  =    f_{\!\ssc\zb} \tilde{X}_{\!\ssc\zc_n} +  \tilde{X}_{\!\ssc\zb_n} f_{\!\ssc\zc}  
   +   i |z|^2  \sum_m \tilde{X}_{\!\ssc\zb_m}  {\tilde{K}}_{\!\ssc\zc_{n\bar{m}}}  \;,
\sea
\tilde{X}_{\!\ssc\zb\zc_{\,\bar{n}}}  \!
=  f_{\!\ssc\zb} \tilde{X}_{\!\ssc\zc_{\bar{n}}}
   +  \tilde{X}_{\!\ssc\zb_{\bar{n}}}  f_{\!\ssc\zc}  
   + i  |z|^2  \sum_m  {\tilde{K}}_{\!\ssc\zb_{m\bar{n}}}  \tilde{X}_{\!\ssc\zc_{\bar{m}}}  \;,
\sea
\tilde{K}_{\!\ssc\zb\zc _{\,m\bar{n}}} \!
   =  \!   f_{\!\ssc\zb} {\tilde{K}}_{\!\ssc\zc_{m\bar{n}}}  \! 
   +  \!  {\tilde{K}}_{\!\ssc\zb_{m\bar{n}}}  f_{\!\ssc\zc}  \!
   +  \!  i|z|^2  \!  \sum_l \! \tilde{K}_{\!\ssc\zb _{l\bar{n}}} \tilde{K}_{\!\ssc\zc _{m\bar{l}}}  \!
   -  \! i  \tilde{X}_{\!\ssc\zb_{\bar{n}}}  \tilde{X}_{\!\ssc\zc_m}  \!
   +   \!\frac{i|z|^2 \tilde{g}_{m\bar n} }{\hbar}
     \sum_l \tilde{X}_{\!\ssc\zb_l} \tilde{X}_{\!\ssc\zc_{\bar{l}}} \;,
\sea
\label{SDp-z}
\eea
and 
\bea &&
f_{\!\ssc\zb\zc}  \!
  = f_{\!\ssc\zb}  f_{\!\ssc\zc}  
   + \hbar  {X}_{\!\ssc\zb_m} g^{m\bar{n}} {X}_{\!\ssc\zc_{\bar{n}}} \;,
\sea
{X}_{\!\ssc\zb\zc_{\,{n}}}  \!
  =  f_{\!\ssc\zb} {X}_{\!\ssc\zc_n} +  {X}_{\!\ssc\zb_n} f_{\!\ssc\zc}  
   +  i \hbar {X}_{\!\ssc\zb_l } g^{l\bar{m}}   {{K}}_{\!\ssc\zc_{n\bar{m}}}  \;,
\sea
{X}_{\!\ssc\zb\zc_{\,\bar{n}}}  \!
=   f_{\!\ssc\zb} {X}_{\!\ssc\zc_{\bar{n}}}
   +  {X}_{\!\ssc\zb_{\bar{n}}}  f_{\!\ssc\zc}  
   + i  \hbar   {{K}}_{\!\ssc\zb_{l\bar{n}}} g^{l\bar{m}}  {X}_{\!\ssc\zc_{\bar{m}}}  \;,
\sea
{K}_{\!\ssc\zb\zc _{\,m\bar{n}}} \!
   =  \!   f_{\!\ssc\zb} {{K}}_{\!\ssc\zc_{m\bar{n}}}  \! 
   +  \!  {{K}}_{\!\ssc\zb_{m\bar{n}}}  f_{\!\ssc\zc}  \!
 + i\hbar   {K}_{\!\ssc\zb _{l\bar{n}}} g^{l\bar{o}} {K}_{\!\ssc\zc _{m\bar{o}}}
   -  \! i  {X}_{\!\ssc\zb_{\bar{n}}}  \tilde{X}_{\!\ssc\zc_m}  \!
  + i g_{m \bar n}  g^{l\bar{o}} X_{\zb l} X_{\zc \bar o} \,.
\sea
\label{SDp-w}
\eea
The above results illustrate what is said above, that the
symmetry data for any two observables have a product
which gives the symmetry data of the product observable
-- almost but not exactly. The elements of
the metric and the inverse metric tensor are involved too.
Upon a more careful thinking, however, that feature is not
unreasonable, in fact it is quite normal. In a non-Euclidean 
space, or even in an Euclidean space depicted in non-Cartesian 
coordinates, even the algebraic relations between classical
observables commonly involve the metric. For example,
the energy of a free Newtonian particle with the momentum 
components $p_{\!\ssc r}$, $p_{\!\ssc\theta}$, 
and $p_{\!\ssc\psi}$ in spherical coordinates is  
$E= \frac{1}{2m} \!\left( p_{\!\ssc r}^2 + \frac{1}{r^2} p_{\!\ssc\theta}^2 
  + \frac{1}{r^2 \sin^2\!\theta} p_{\!\ssc\psi}^2 \right)$. 
Generally, if the physical space is a curved manifold with metric
$g_{ab}$, we have $E= \frac{1}{2m} g_{ab} p^a p^b$. The metric
is in fact always there, only that for the case of an Euclidean
geometry in the standard Cartesian coordinates, its values
are simply $\delta_{ab}$ at every point and hence we can 
write the expression without showing it explicitly. Here for
calculating the symmetry data for any operator product
through the $f_{\!\ssc\zb}$ functions, however, we always
need the values of the metric and inverse metric tensor
elements at the phase space point. The truth is that the 
metric tensor elements are also all over Eqs.(\ref{SDp-h}) 
and (\ref{SDp-z}) implicitly, as in much the same form as 
they explicitly show up in Eq.(\ref{SDp-w}). The other apparent 
problem is that the expressions in terms of the $z^n$ coordinates
also depend on $|z|^2$, but it we can easily avoided by 
looking only at the normalized $\left|\phi\rra$. In our 
formulation with $z^n$ bearing the physical dimension of 
length, the preferred normalization is actually $|z|^2 = 2\hbar$ 
at which $f_{\!\ssc\zb}=H_{\!\ssc\zb}$ \cite{078}.

The next thing to examine is if the symmetry data for an 
observable on a fixed state $[\phi_{\ssc 0}]$ can be determined 
experimentally, at least in principle. We consider first a
Hermitian $\zb$ and the $H_{\!\ssc\zb}$ functions first. 
For a normalized state, $H_{\!\ssc\zb}([\phi_{\ssc 0}])$ (and 
$f_{\!\ssc\zb}([\phi_{\ssc 0}])$) is just the familiar expectation 
value of an operator $\zb$. $\widetilde{K}_{\!\ssc\zb_{m\bar{n}}}$ 
are the matrix elements on a supposedly known set of orthonormal
 states, hence have no problem either. With those, determining
$\widetilde{X}_{\!\ssc\zb_n} ([\phi_{\ssc 0}])$ means
determining $\widetilde{X}_{\!\ssc\zb_n} (\phi_{\ssc 0})$
for the normalized $\left|\phi_{\ssc 0}\rra$ up to an overall 
phase factor as the one among the $z^n$ coordinates. 
The phase is certainly not to be determined. More 
specifically, if we take as the basis $\left|z_n\rra$ the 
set of eigenstates for $\zb$, assuming no degeneracy, the
nonzero $\widetilde{K}_{\!\ssc\zb_{m\bar{n}}}$ are exactly 
$\widetilde{K}_{\!\ssc\zb_{n\bar{n}}} = \frac{-i\lambda_n}{2\hbar} $
with the eigenvalues $\lambda_n$ and we simply have
$\widetilde{X}_{\!\ssc\zb_n}= \frac{i\lambda_n}{2\hbar}  \bar{z}^n$
(no sum). Gone are the days when people debate if
the physical quantum state can be determined or observed.
On the one hand, we have theoretical analyses, most notably
the line of work presented in Ref.\cite{Sk}, illustrating 
the involved mathematics; on the other there has been 
developed experimental efforts in quantum optics \cite{L}, 
especially the technique of optical homodyne tomography
 \cite{SBRF}, to actually measure the state. Of course, any 
practical measurement gives only approximations. Then, 
the $z^n$ coordinates for a normalized 
$\left|\phi_{\ssc 0}\rra$, up to an overall 
phase factor, and  $\widetilde{X}_{\!\ssc\zb_n}$ 
(and $\widetilde{X}_{\!\ssc\zb_{\bar{n}}}$ as the complex 
conjugate) can be determined in principle. For non-Hermitian 
$\zb$, we can take it as a `complex function' with real 
and imaginary parts as Hermitian operators
(see the above footnote). With the linear structure of
the algebra well-preserved in the K\"ahler product algebra
and the symmetry data algebra, it is then a trivial extension
to incorporate the case of such a  non-Hermitian $\zb$.
The symmetry data in terms of an $f_{\!\ssc\zb}$ function, 
given in Eqs.(\ref{SD-z}) and (\ref{SD-w}), then clearly 
pose no further qualitative difficulty. 
$\tilde{X}_{\!\ssc\zb_n} ([\phi_{\ssc 0}])$ and
$\tilde{X}_{\!\ssc\zb_{\bar{n}}} ([\phi_{\ssc 0}])$ or
$\tilde{X}_{\!\ssc\zb_n} ([\left|\phi_{\ssc 0}\rra])$  and
$\tilde{X}_{\!\ssc\zb_{\bar{n}}} ([\left|\phi_{\ssc 0}\rra])$,
up to the overall phase factor in the $z^n$ coordinates, 
as well as $\tilde{K}_{\!\ssc\zb_{m\bar{n}}}([\phi_{\ssc 0}])$, 
${X}_{\!\ssc\zb_n} ([\phi_{\ssc 0}])$, 
${X}_{\!\ssc\zb_{\bar{n}}} ([\phi_{\ssc 0}])$, and
${K}_{\!\ssc\zb_{m\bar{n}}}([\phi_{\ssc 0}])$ can all 
be determined likewise. The kind of expression like
${X}_{\!\ssc\zb_n} ([\phi_{\ssc 0}])$ means only the
complex number of ${X}_{\!\ssc\zb_n}$ after the
fixed values of the coordinates, $w^n$ or $z^n$ 
are fixed for the ray of the state vector 
$\left|\phi_{\ssc 0}\rra$. Note that the values of
$\tilde{K}_{\!\ssc\zb_{m\bar{n}}}$, ${X}_{\!\ssc\zb_n}$, 
${X}_{\!\ssc\zb_{\bar{n}}}$, and ${K}_{\!\ssc\zb_{m\bar{n}}}$
are in general  independent of the undetermined phase
factor, also in $w^n$ coordinates. Moreover, knowing the 
coordinates, we have all the elements of the metric tensors 
at the state, which can be considered as a theoretical input 
at this point. From the perspective of Refs.\cite{066,070}, 
the metric gives a real number distance between the two 
quantum particle states, the two points in the quantum model 
of the physical space, and its connection to quantum observables 
is explored in Ref.\cite{078}. One would like to find a way 
to measure or verify the metric tensor experimentally but
we will not go more into that here. It is important to note the 
following. While an $f_{\!\ssc\zb}$ function, unlike the 
$H_{\!\ssc\zb}$ function, has nontrivial higher derivatives,
they have however definite relation to the first and second 
order ones through the metric \cite{Sd} as a result of the 
K\"ahlerian nature of the $f_{\!\ssc\zb}$ function. That goes 
for both the descriptions in terms of the $z$- or $w$-coordinates. 
Hence ${X}_{\!\ssc\zb_n}$ , ${X}_{\!\ssc\zb_{\bar{n}}}$, 
and ${K}_{\!\ssc\zb_{m\bar{n}}}$, or $\tilde{X}_{\!\ssc\zb_n}$ , 
$\tilde{X}_{\!\ssc\zb_{\bar{n}}}$, $\tilde{K}_{\!\ssc\zb_{m\bar{n}}}$, 
make up the complete set of independent derivatives, or 
covariant derivatives. The fact that the operator, or its 
representation $f_{\!\ssc\zb}$ or $H_{\!\ssc\zb}$, can 
indeed be determined, at least locally, by the values of all 
its derivatives at a point  is simply the mathematics of the 
Taylor series expansion.
 
\section{Physics Discussions}
We have argued above that the symmetry data for an 
operator at a fixed physical state can in principle be 
experimentally determined. Note that it is actually
most directly represented by the set of complex numbers 
$\{ f_{\!\ssc\zb}([\phi_{\ssc 0}]),   {X}_{\!\ssc\zb_n} ([\phi_{\ssc 0}]), 
 {X}_{\!\ssc\zb_{\bar{n}}} ([\phi_{\ssc 0}]), 
 {K}_{\!\ssc\zb_{m\bar{n}}}([\phi_{\ssc 0}]) \}$.
The corresponding set in terms of  the homogeneous
coordinates $z^n$, and even the representation through
the $H_{\!\ssc\zb}$ function and its derivatives, have
the complication of having to be restricted to the normalized
state vectors, with an unphysical overall phase factor for
the $z^n$ coordinates formally involved in the first 
derivatives. That does no harm. In fact, we can even take 
a conventional definition of the coordinates to fix it,  for 
example  taking $z^{\!\ssc 0}$ as real. Unlike the $w^n$ 
coordinates, the $z^n$ coordinates cover the whole space 
of physical states, $\mathcal{P}$.  All that, however, is only 
of a theoretical interest. For the idea of the noncommutative
(number) value of an observable to be of any practical use,
one does not want to determine them as indirectly as  
discussed above. We want their direct measurements. 
We argue below that such an idea, although going
much beyond the familiar, may not be as crazy as it may 
sound to those who never thought about it before.

Let us look into the notion of a measurement of physical
quantities very carefully. A measurement is a controlled
interaction with the physical system we are interested in 
with a goal of extracting a particular information about
its state before that interaction. It is actually not necessary 
to worry about the possible change of the state during the 
process. To actually obtain the information or the `values'
we are after may, however, be a complicated issue. In 
general, we need to have a good (theoretical) understanding 
of the physics of the measuring process and perform some 
calculations. The world is quantum and the information
about physical system is intrinsically quantum in nature, 
though it may be approximated by, or in many cases degraded
to, the classical information represented by a few real numbers.
Extracting a piece of quantum information from a system 
should really be taken as a kind of measuring process, and 
it is certainly not the one giving the `real number answers'. 
For the direct measurements in which the answer is read 
out of an apparatus, the key is to have the right apparatus 
and a good calibration of the output scale. We essentially 
measure by comparison. We compare the ``value" we measure, 
as indicated on the reading scale, with ``known value" of the 
quantities which may be a conventionally chosen standard 
unit. The comparison itself never gives us the real number 
readings though. We put that in ourselves. The truth is that 
nothing in the nature ever points to the idea of the physical 
quantities being real valued. In all (classical) measurements,
it is our calibration of the output reading scale in the 
measuring device that enforces the reading as a real number,
and even that can only be taken as an approximate range. 
The bottom line is that the real numbers as the representation 
of the values of physical quantities are nothing more than 
a mathematical model we use to describe nature. But the 
model shows inadequacy in the quantum regime.
We are interested in finding the better way to model those 
`values'. To consider the full practical implementations 
of what is here theoretically discussed, one may have to 
devise an appropriate apparatus and calibrate its output 
with the noncommutative values. 
 
Taking a step backward to the conservative consideration of the
von Neumann measurements on a quantum observable, one 
must note that a single real eigenvalue outcome in itself hardly 
gives any information about the ``value" of the observable on 
a particular state. The best real number representation of the 
value is the expectation value as the mean of a sufficient 
statistical collection of the eigenvalue outcomes from repeated 
measurements. But then it has a standard statistical uncertainty. 
However, we have more than the expectation value and the 
uncertainty. We have the whole distribution which contains 
its full information in terms of an infinite number of real 
numbers in all the moments. Depending on the desired
precision, we can take enough statistics and calculate a large 
enough number of the lower moments to get to an approximation 
of the full information content for the value of the observable 
on the fixed state. There is no reason, other than ignorance, 
to discard all those real number information in all the moments 
and use only one or two as done in a standard discussion
about quantum observables. The full infinite  set of real 
numbers, or the whole distribution, should be taken as the 
value of the observable on a state. The latter has an obvious 
parallel with the notion of the symmetry data. Again, the value
of a quantum observable for a state should be represented by 
the element of a noncommutative algebra rather than by the 
real numbers, \emph{i.e.} of a commutative one.  Such an 
element can also be represented by an infinite set of the 
real/complex numbers.

The representation of our noncommutative value for an
observable in terms of the symmetry data as the infinite
set of complex numbers has the first number being the
functional value of $f_{\ssc \zb}$, exactly the expectation 
value. It is real for a Hermitian $\zb$, and is no doubt
the best single real number representation of the value
of $\zb$ as a physical quantity.  What the full 
noncommutative value represents is the full information
the theory can give mathematically matching any 
observable to a fixed physical state. It goes beyond 
determining the whole statistical distribution of results
from von Neumann measurements. We have seen above
that the set of complex numbers essentially contains 
enough information to determined the exact physical 
state as well as all matrix elements of the observable 
with respect to the fixed coordinate basis. And the
noncommutative value for $\zb$ and $\zc$ allows
us to determine that for the products $\zb\zc$ and
$\zc\zb$, satisfying whatever commutation relation
they have.  

Intuitively, when the state is fixed, all physical properties should 
be fixed. To have the idea implemented in quantum mechanics
is workable, we simply have to go beyond the idea of describing 
each simple physical property, such as a position coordinate, 
by a single real number. In our opinion, the stubborn physicists' 
attachment  to the real number values is what makes quantum 
theory seemingly counter-intuitive. We can and should go 
beyond that. There is no hidden variables in quantum mechanics
and the noncommutative variables of position and momentum
fully describe the states, but the full description of their values
for a definite state has been `hidden' beyond  any single
real number representation for each variable. The latter 
gives an incomplete description, while the complete 
information is always there in the noncommutative values.
The mathematician Takesaki talked about operator 
algebras as `a number theory in analysis'  \cite{T}.  From a
physics point of view, observables are represented by the 
operators, the variables to be evaluated on a physical state. 
The notion of their noncommutative values may be really be 
seen as noncommutative numbers. A proper representation 
of a $C^*$-algebra, which is what physicists should focus on 
as a candidate for the observable algebra \cite{St}, gives 
an operator algebra on a Hilbert space. The corresponding 
projective Hilbert space is a space of pure states which is 
a mathematical object dual to (the representation of) the 
$C^*$-algebra \cite{S,AS}.  $C^*$-algebras are noncommutative 
geometric objects \cite{C}. A  $C^*$-algebra of the quantum 
operators, as functions of the six position and momentum 
operators \cite{070}, may be taken as having them as 
coordinate observables of the geometry whose 
noncommutative values for each physical state may 
plausibly be seen as an alternative description of an infinite 
number of real/complex number coordinates of the 
projective Hilbert space.  The study of that idea is reported 
in a companion paper  Ref.\cite{078}.

  \vspace*{.2in}
\noindent{\bf Acknowledgments \ }
We thank Suzana Bedi\'c for helping to improve the 
language of the presentation.
The authors are partially supported by research grants 
number 109-2112-M-008-016 and 107-2119-M-008-011 of the MOST of Taiwan.


\begin{thebibliography}{99}
\bibitem{078}
Kong O.C.W., Liu W.-Y., Noncommutative Coordinate Picture of the Quantum Physical/Phase Space,
arXiv: 1903.11962, NCU-HEP-k078, and references therein.
\bibitem{066}
Chew C.S., Kong O.C.W., Payne J.,  
 A Quantum Space behind Simple Quantum Mechanics, 
Adv. High Energy Phys. {2017} 
 (2017) 4395918
\bibitem{070}
Chew C.S., Kong O.C.W., Payne J., Observables and Dynamics, Quantum to Classical from a Relativity Symmetry and Noncommutative-Geometric Perspective,  
J. High Energy Phys. Grav. Cosmo. {5} (2019) 553-586.
\bibitem{081}
 Kong O.C.W.,  A Geometric Picture of Quantum 
Mechanics with Noncommutative Values for Observables, Results in Phys.
{\bf 19} (2020) 103636. 
\bibitem{Sd} 
Schilling T.A.,  Geometry of Quantum Mechanics, Ph.D dissertation, the Pennsylvania
State University, 1996;
 Ashtekar A., Schilling T.A., Geometrical Formulation of Quantum Mechanics, {On Einstein's Path},
 Harvey A.(ed.), Springer, New York, 1998, p. 23-65, [gr-qc/9706069]. 
\bibitem{Cps}
Cirelli R., Mani\`a A., Pizzocchero L., 
Quantum Mechanics as an Infinite-Dimensional Hamiltonian System with Uncertainty Structure: Part I, 
J. Math. Phys. {31} (1990) 2891-2897.
\bibitem{Sk}
Schroeck F. E. Jr.,  {Quantum Mechanics on Phase Space},
Kluwer Academic Publishers, Netherlands, 1996.
\bibitem{L}
Leonhardt U., {Measuring the Quantum State of Light}, Cambridge University Press, Cambridge, 1997.
\bibitem{SBRF}
Smithey D.T.,   Beck M., Raymer M.G., Faridani  A., 
Measurement of the Wigner Distribution and the Density Matrix of a Light Mode Using Optical Homodyne Tomography: Application to Squeezed States and the Vacuum, 
Phys. Rev. Lett. {70} (1993) 1244-1247.
\bibitem{T}
Takesaki M.,  {Theory of Operator Algebras II}, Springer, Berlin, 2003.
\bibitem{St} 
Strocchi F.,  {An Introduction to the Mathematical Structure of Quantum Mechanics}, World Scientific, 2008.
\bibitem{S}
Shultz F.W., Pure States as a Dual Object for $C^*$-Algebras,  Commun. Math. Phys. {82}
(1982) 497-509.
\bibitem{AS}
Alfsen E.M., Shultz F.W., {State Spaces of Operator Algebras}, 
Birkh\"auser, Boston, 2001.
\bibitem{C}
Connes A., {Noncommutative Geometry}, Academic Press, 1994.
\end{thebibliography}
\end{document}